# Leveraging Sensory Data in Estimating Transformer Lifetime

Mohsen Mahoor, Alireza Majzoobi, Zohreh S. Hosseini, and Amin Khodaei
Dept. of Electrical and Computer Engineering
University of Denver
Denver, CO, USA
Mohsen.Mahoor@du.edu, Alireza.Majzoobi@du.edu, Zohreh.Hosseini@du.edu, Amin.Khodaei@du.edu

*Abstract*—Transformer lifetime assessments plays a vital role in reliable operation of power systems. In this paper, leveraging sensory data, an approach in estimating transformer lifetime is presented. The winding hottest-spot temperature, which is the pivotal driver that impacts transformer aging, is measured hourly via a temperature sensor, then transformer loss of life is calculated based on the IEEE Std. C57.91-2011. A Cumulative Moving Average (CMA) model is subsequently applied to the data stream of the transformer loss of life to provide hourly estimates until convergence. Numerical examples demonstrate the effectiveness of the proposed approach for the transformer lifetime estimation, and explores its efficiency and practical merits.

## Nomenclature

*Parameters:*

$F_{AA}$    Aging acceleration factor of insulation.
$F_{AA,n}$    Aging acceleration factor for the temperature during the time interval $\Delta t_n$.
$F_{EQA}$    Equivalent aging factor for the total time period.
$K$    Ratio of actual load to rated load.
$m/n$    An empirically derived exponent used to calculate the variation of $\Delta\Theta_H/\Delta\Theta_{TO}$ with changes in load.
$R$    Ratio of load loss at rated load to no-load loss on the selected tap position.
$\Delta t_n$    Time interval.
$\theta_H$    Winding hottest-spot temperature (ºC).
$\theta_A$    Average ambient temperature during the load cycle (ºC).
$\Delta\theta_H$    Average ambient temperature during the load cycle (ºC).
$\Delta\theta_{TO}$    Top-oil rise over ambient temperature (ºC).

*Subscripts:*

$H$    Hottest-spot
$i/U$    Initial/Ultimate values
$R$    Rated value
$TO$    Top oil
$w$    Winding

## I. Introduction

ASSET MANAGEMENT has always been an important responsibility of utility companies to ensure system health and to prevent undesired component failures through timely upgrade and upkeep, and as a result, deliver the best service to electricity customers and reduce the power system outages as much as possible. This concept has always been central to utility operations, however, given the aging power system infrastructure in the U.S., mainly built in 1950s and 60s, and increasing customers' expectations on the quality of service, this topic is now more important than ever [1]-[5].

From asset management perspective, the transformer asset management is one of the most significant and critical subjects for electric utility companies. Transformers play a key role in power system reliability as their failure lead to large area power outages. In addition, repairing and changing the transformer is usually time-consuming and costly for electric utility companies [1], [2]. Therefore, it is critical and important for utility companies to predict the transformer lifetime in order to plan for maintenance or replacement, if necessary. The transformer lifetime mainly depends on its insulation condition due to higher failure probability of insulation compared with other components of transformer. The end of transformer life is a condition which the insulation can no longer withstand the stresses over fault conditions. Moreover, amongst different aging factors of transformer insulation, such as insulation moisture, oxygen amount, internal temperature, and electrical stress, aging of a transformer highly depends on the thermal degradation of the insulation during its normal life cycle. Thus, the primary aging factor is transformer's internal temperature, specifically at the hottest-spot, which is a function of transformer's loading and ambient temperature [6]-[8].

In [5], a two-stage maintenance scheduler (mid-term and short-term) is presented for transformer asset maintenance management. The effect of temperature and electrical stress on transformer insulation is experimentally analyzed in [9], through measuring and diagnosing the main characteristics of insulation, such as the tan δ and resistivity. IEEE presents a set of experimental equations in the IEEE Std. C57.91-2011, "Guide for Loading Mineral-Oil-Immersed Transformers", for calculation of power and distribution transformers loss of life [10]. This standard, which was issued in 1981 for the first time and has been revised several times, has always been a main reference of industry and researchers for calculation of the transformer loss of life [7], [11]-[13]. The study in [7] estimates time to failure of the transformer insulation, based on this IEEE standard and the forecasted data for load profile and ambient temperature. Various machine learning methods are applied in [11] to estimate transformer loss of life based on this IEEE standard. In [12], the reliability of the IEEE standard thermal model is enhanced by applying the data-quality control and data-set screening procedures. In [13], the thermal model of the above-mentioned standard is modeled via MATLAB Simulink in order to estimate transformer loss of life based on its hottest-spot temperature. In addition, the effect of transformer cooling methods on reduction of its loss of life is investigated. A parametric lifetime model is presented in [14] to find the lifetime distribution of individual transformers and is extended to provide lifetime prediction for the overall fleet of transformers. In [7] and [15], a smart charging framework is proposed to mitigate the effects of electric vehicles on distribution assets, such as transformers. Various types of Artificial Neural Networks (ANN) are also utilized for transformer lifetime estimation [16]-[18]. In [16] fuzzy modeling is utilized for transformer asset management and increasing the remaining life of transformer. The intelligent system with the capability of comprehensive detection and diagnosis of failure modes in the active part of power transformers, via online monitoring system, is proposed in [17]. The authors of [18] propose a simpler and more accurate alternative model to IEEE standard for calculation of transformer internal temperature of 25 kVA mineral-oil-immersed transformers, and accordingly calculating transformer loss of life.

The primary objective of this paper is to provide a sensory model framework to measure the transformer internal temperature, i.e., the winding hottest-spot temperature, plug these measured values into the IEEE Std. C57.91-2011 to calculate the transformer loss of life at each time interval, and accordingly determine a good estimate of transformer lifetime. A Cumulative Moving Average (CMA) model is applied to the data stream of the transformer loss of life for this purpose. Using the CMA value, transformer lifetime is estimated at each time interval up until it is converged. Numerical examples, to be carried out in this paper, justify that the transformer lifetime can be estimated using the measured sensory data of the winding hottest-spot temperature and the proposed CMA model.

The rest of the paper is organized as follows. Section II presents the IEEE standard formulations, introduces the sensory model structure for a distribution transformer, and proposes the CMA model to estimate the transformer lifetime. Section III presents numerical analyses to show the effectiveness of the proposed model. Conclusions are presented in Section IV.

II. TRANSFORMER MODEL OUTLINE AND FORMULATION

A sensory data in line with CMA approach are employed in a dynamic manner to estimate transformer lifetime. In what follows, first, a sequence of nonlinear and exponential functions based on the IEEE Std.C57.91-2011 is presented to calculate transformer loss of life. Then, a sensory model structure for measuring transformer winding-hottest-spot temperature is introduced. Finally, CMA model is proposed in order to apply to the data stream of transformer loss of life, and consequently estimate transformer lifetime.

*A. The IEEE Standard Model*

As mentioned, the internal temperature of the transformer, which is a function of transformer loading and ambient temperature, is the primary factor on the aging of the transformer insulation. The IEEE Std. C57.91-2011 provides a model for calculation of the transformer loss of life based on the winding hottest-spot temperature. As the temperature does not have a uniform distribution in the transformer, the hottest-spot is considered in calculations. The Arrhenius' chemical reaction rate theory is the source of the IEEE standard experimental equations for calculation of transformer loss of life. Equation (1) defines the per unit life of transformers,

$$\text{Per unit life} = A \exp\left(\frac{B}{\theta_H + 273}\right), \qquad (1)$$

where $A$ is a modified per unit constant and $B$ is the aging rate. $A$ is equal to $9.8 \times 10^{-18}$ which is calculated based on selection of 110 °C as the temperature for "one per unit life" and $B$ is computed between 11350 and 18000 in various experiments; a value of 15000 is chosen for $B$ in IEEE Std. C57.91-2011.

Substitution of constants $A$ and $B$ in (1), gives Aging Acceleration Factor (AAF) for a given winding hottest-spot temperature (2).

$$F_{AA} = \exp\left(\frac{15000}{383} - \frac{15000}{\theta_H + 273}\right). \qquad (2)$$

The hottest-spot temperature on the winding is a critical point as in this temperature transformer insulation degrades. As (2) demonstrates, the insulation's lifetime and accordingly the transformer's lifetime is exponentially related to hottest-spot winding temperature. At 110°C, AAF equals 1 which means transformer will have its normal life expectancy. While, for hottest-spot winding temperature higher/lower than 110°C the transformer lifetime decreases/extends. It is worth to mention that the phrase "loss of life" commonly means "loss of insulation life", although "insulation" is frequently omitted.

The equivalent aging of the transformer in a desired time period is obtained based on (2), as follows:

$$F_{EQA} = \sum_{n=1}^{N} F_{AA_n} \Delta t_n \bigg/ \sum_{n=1}^{N} \Delta t_n , \qquad (3)$$

where $\Delta t_n$ is time interval, $n$ is the time interval index and $N$ is the total number of time intervals. The insulation loss of life is accordingly calculated as below:

$$LOL(\%) = \frac{F_{EQA} \times t \times 100}{\text{Normal insulation life}}, \qquad (4)$$

The IEEE standard considers 180000 hours as the normal insulation lifetime for distribution transformers to be included in (4).

As (1)-(4) show, the first step for calculation of transformer loss of life is computing hottest-spot temperature (5).

$$\theta_H = \theta_A + \Delta\theta_{TO} + \Delta\theta_H , \qquad (5)$$

In this equation, $\theta_A$ represents ambient temperature, $\Delta\theta_{TO}$ is top-oil rise over ambient temperature which is calculated by (6), and $\Delta\theta_H$ is the winding hottest-spot rise over top-oil temperature, calculated by (7).

$$\Delta\theta_{TO} = (\Delta\theta_{TO,U} - \Delta\theta_{TO,i})(1 - \exp(-\frac{1}{\tau_{TO}})) + \Delta\theta_{TO,i} , \qquad (6)$$

$$\Delta\theta_H = (\Delta\theta_{H,U} - \Delta\theta_{H,i})(1 - \exp(-\frac{t}{\tau_w})) + \Delta\theta_{H,i} . \qquad (7)$$

Furthermore, the initial and ultimate values of $\Delta\theta_{TO}$ and $\Delta\theta_H$ in (6) and (7) are calculated through (8)-(11), as follows:

$$\Delta\theta_{TO,i} = \Delta\theta_{TO,R} (\frac{K_i^2 R + 1}{R + 1})^n , \qquad (8)$$

$$\Delta\theta_{TO,U} = \Delta\theta_{TO,R} (\frac{K_U^2 R + 1}{R + 1})^n , \qquad (9)$$

$$\Delta\theta_{H,i} = \Delta\theta_{H,R} K_i^{2m} , \qquad (10)$$

$$\Delta\theta_{H,U} = \Delta\theta_{H,R} K_U^{2m} . \qquad (11)$$

Note that $\Delta\theta_{TO,R}$, $\Delta\theta_{H,R}$, $R$, $m$, $n$ are constants and come from transformer characteristics. $m$ and $n$ depend on the transformer cooling system and vary between 0.8 and 1 [10, Table 4].

Considering (1)-(11), derived from the mentioned IEEE standard, it can be seen that the transformer winding hottest-spot, i.e., $\theta_H$, is the main factor in calculating the transformer loss of life at each time interval. Moreover, the value of the winding hottest-spot temperature is governed by ambient temperature, initial value of transformer load ratio, and ultimate value of transformer load ratio at each time interval, i.e., $\theta_A$, $K_i$ and $K_U$, respectively. Nevertheless, by having a temperature sensor and measuring the transformer winding hottest-spot temperature, the transformer loss of life could be calculated at each time interval.

### B. Sensory Model Structure for Transformer

This section develops a sensory model structure to measure the winding hottest-spot temperature via a temperature sensor and consequently calculates the transformer loss of life based on the above-mentioned equations. This temperature sensor takes the responsibility of measuring the real temperature values of the transformer winding hottest-spot at each time interval. Next, these measured values are utilized to calculate the transformer loss of life. In other words, the value of the transformer loss of life is updated at each time interval using the sensory measured value of the winding hottest-spot temperature. As mentioned in the IEEE Standard, normal lifetime of distribution transformers is 180000 hours, i.e., 20 years. However, this value for normal lifetime is not a fixed number during transformer operational lifecycle, and could be shortened or prolonged in line with the variation of the winding hottest-spot temperature. In other words, if the winding hottest-spot temperature increases or decreases, the transformer loss of life would be increased/decreased, and consequently the normal lifetime for the transformer would be reduced/extended.

Employing the discussed equations and the measured data from the temperature sensor, the transformer loss of life is calculated at each time interval, then the transformer lifetime is estimated in a dynamic manner. Fig. 1 shows the overall sensory model structure to estimate the transformer lifetime. As shown in this figure, transformer lifetime, which could possess a different value from the normal lifetime of the transformer, could be estimated using the sensory data captured from the transformer winding hottest-spot temperature.

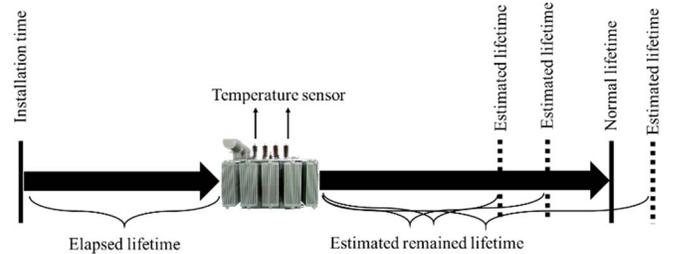

Fig.1. Sensory model structure for transformer lifetime using temperature sensor.

### C. Cumulative Moving Average Model to Estimate Transformer Lifetime

In order to estimate transformer lifetime, a CMA model is applied to the recorded data stream, generated form the values of the transformer loss of life. In this regard, the calculated values for the transformer loss of life arrive in an ordered data stream, and the CMA model apply the averaging operator to all the ordered data values up until the current point. Through averaging, the model take the advantages of all the calculated data for the loss of life to estimate the transformer lifetime. Using the CMA model, as each new data point arrives, the average value for the transformer loss of life at the time of the measuring the transformer winding hottest-spot temperature is calculated for all of the ordered values up to that current point, and the lifetime is accordingly updated. Equation (12) demonstrates the CMA model for the ordered data values of the transformer loss of life.

$$\text{CMA}_n (\%) = \frac{(LOL_1 + LOL_2 + ... + LOL_n)}{n}, \quad (12)$$

where, $LOL_1, LOL_2,..., LOL_n$ represent the ordered data stream for the transformer loss of life, $n$ is the number the data stream arrived to the model, and $CMA_n$ represents the CMA value for the ordered data stream of the transformer loss of life. Using, (13) the cumulative average is dynamically updated when a new value for the transformer loss of life, i.e., $LOL_{n+1}$, becomes available.

$$\text{CMA}_{n+1}(\%) = \frac{LOL_{n+1} + n \times CMA_n}{n+1}, \quad (13)$$

As the CMA model is updating the value of the transformer loss of life at each time interval, the transformer lifetime is estimated at that corresponding time interval. Equation (14) is used to calculate the estimated transformer lifetime in a desired time interval.

$$\text{Estimated Lifetime}_n = \frac{\Delta t_n}{8760 \times CMA_n} + \frac{n \times \Delta t_n}{8760}. \quad (14)$$

The first term in (14) is the estimated remained lifetime using the CMA value of the transformer loss of life at that time interval. The second term represents the elapsed lifetime for the transformer during the period of feeding the temperature sensor data points into the estimating process. This estimating process is occurring dynamically up until the value for the transformer lifetime is converged. Fig. 2 depicts the flowchart of the proposed framework for estimating the transformer lifetime in which the sensory data of the transformer winding hottest-spot temperature, formulations of the IEEE standards, and the CMA model are coming together to aim this goal.

III. NUMERICAL EXAMPLES

In order to evaluate the performance of the proposed framework for estimating transformer lifetime, an hourly sequence of data for the transformer winding hottest-spot temperature is synthesized under various weather conditions and transformer's loading. In this respect, an hourly ambient temperature, borrowed from [19], and initial and ultimate values of the transformer load ratio are used. The process of the data synthesis needs some characteristics of the transformer which are borrowed from [6] and tabulated in Table I. Furthermore, a time interval, i.e., $\Delta t_n$, of 1 hour is considered for data synthesis and modeling. A total number of 8760 time intervals are considered, equal to the number of hours in one year.

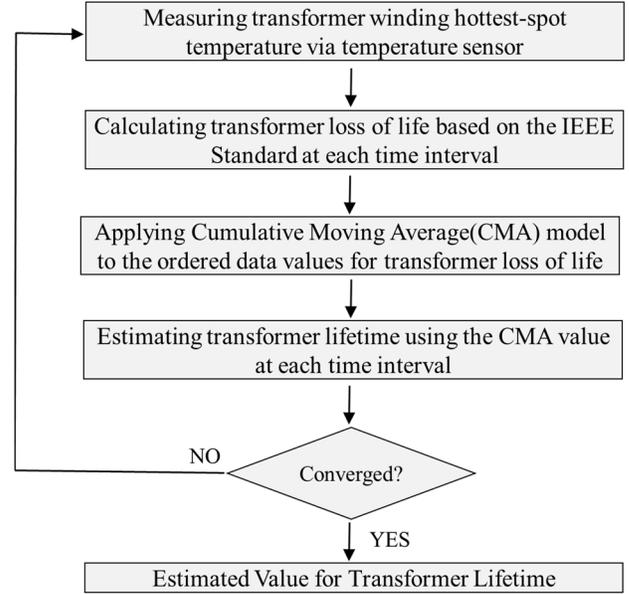

Fig. 2. Flowchart of the proposed framework for estimating transformer lifetime using sensory data and CMA model.

TABLE I
CHARACTERISTICS OF THE STUDIED TRANSFORMER [6]

| $I_{rating}$ | R | m,n | $\Delta\theta_{H,R}$ | $\Delta\theta_{TO,R}$ | $\tau_{TO,R}$ |
|---|---|---|---|---|---|
| 934 A | 7.43 | 0.8 | 17.6 °C | 53.9 °C | 6.8 h |

According to various weather conditions, which imply different ambient temperature, and the load ratio for the transformer, following cases are studied:

Case 1: Mild weather condition
Case 2: Warm weather condition
Case 3: Warm weather conditions along with overloading

**Case 1:** In this case, the transformer is considered to be in a specific place which has a mild climate. This mild weather condition causes the transformer to experience both normal ambient temperature and normal load ratio during operation. The transformer winding hottest-spot temperature is measured via the temperature sensor at each hour, then employed in the proposed framework to estimate the transformer lifetime. Fig. 3 compares the hourly and the CMA values of the transformer loss of life. The convergence process for estimating the value of the transformer lifetime is shown in Fig. 4. As shown in these two figures, due to the fact that the data in the beginning of the measurement horizon are sparse, the CMA value does not represent the transformer loss of life precisely so that the estimated lifetime for the transformer is oscillating. After measuring 8003 sample points, i.e., after 8003 hours, the CMA value is rich enough to be generalized to all the pervious measured data value, and as a result the transformer lifetime value is converged to a constant. The lifetime value for the transformer is estimated to be 37.3 years in this case, much greater than the initially estimated lifetime of 20 years. It should be mentioned that in all the figures, logarithmic scale is considered for Y-axis.

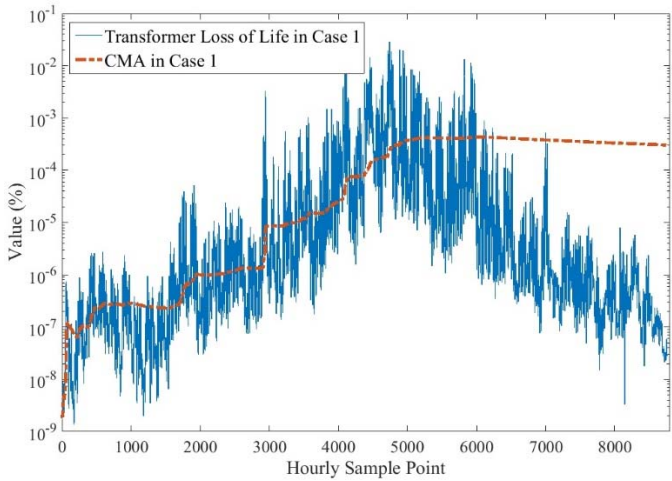
Fig. 3. Hourly and CMA values for transformer loss of life in Case 1.

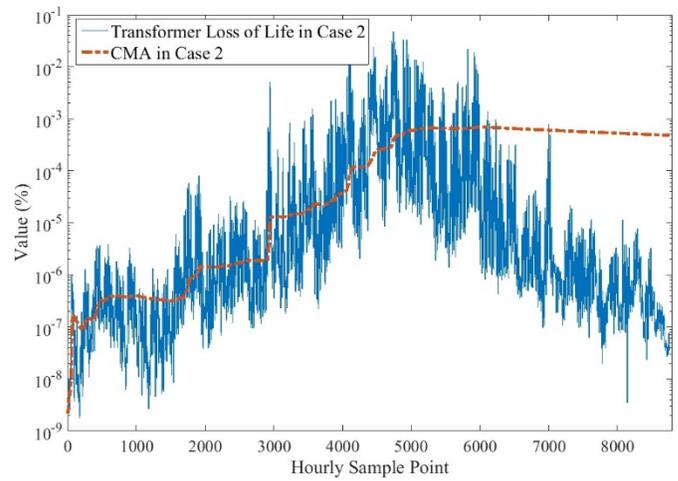
Fig. 5. Hourly and CMA values for transformer loss of life in Case 2.

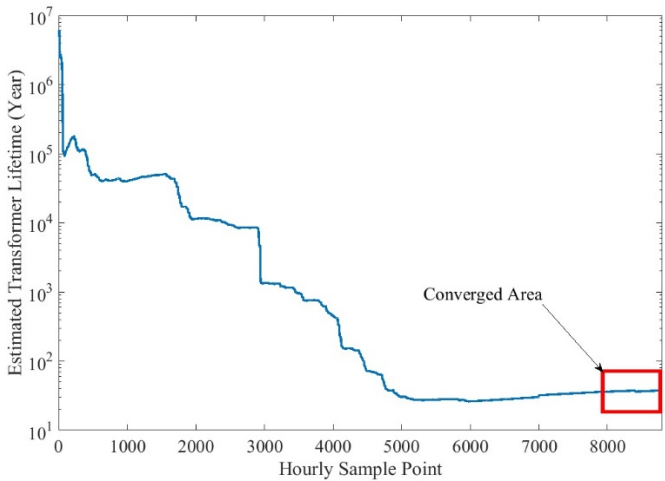
Fig. 4. Estimating transformer lifetime up to its convergence in Case 1.

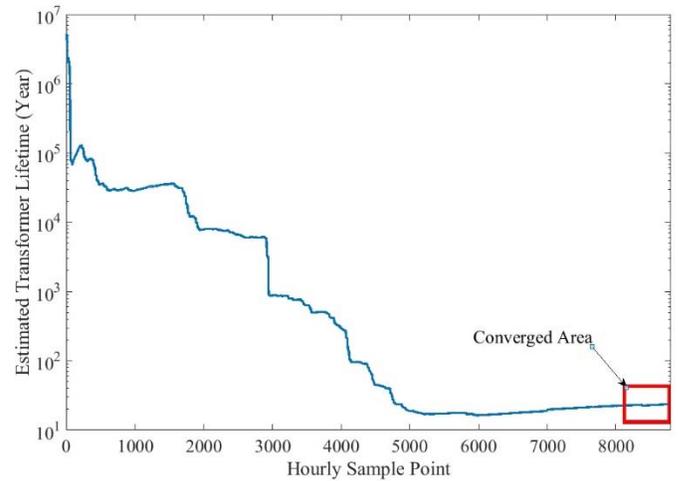
Fig. 6. Estimating transformer lifetime up to its convergence in Case 2.

**Case 2:** A warm weather condition is considered for the transformer in this case. It is clear that there is a correlation between ambient temperature and the transformer load ratio; warm ambient temperature causes higher load ratio to the transformer during operation. Accordingly, the transformer winding hottest-spot temperature will increase. Nevertheless, in order to estimate the transformer lifetime, the data for the winding hottest-spot temperature, measured hourly by the temperature sensor, are fed to the proposed model. Fig. 5 shows the hourly and the CMA values of the transformer loss of life, and Fig. 6 depicts the convergence process for estimating the value of the transformer lifetime. As shown in these figures, after 8150 samples of data stream, the transformer lifetime is converged to 23.5 years, which again is greater than the initially assumed lifetime. The transformer in this case has a shorter lifetime, compared to Case 1, conceivably due to higher temperature at the hottest-spot. Thus, transformer lifetime in the warm climate considerably declines, compared to the mild climate in Case 1, due to the double impact of warm ambient temperature and excessive transformer load ratio on its winding hottest-spot temperature.

**Case 3:** Overloading has a negative impact on the transformer lifetime. Transformer overloading sets the stage for a sharp decline in its lifetime. The more transformer undergoes overloading conditions, the more its winding hottest-spot temperature increases, and the less its lifetime will be. This case investigates the effect of overloading on the transformer lifetime. In this regard, the transformer in Case 2 is assumed to undergo 20% overloading at 3 hours of 20 randomly selected days in a whole year. Fig. 7 compares the estimated transformer lifetime for all these three cases. In Case 3, the proposed framework uses 8340 hourly sample points to estimate the transformer lifetime, and the transformer lifetime is estimated to be 21.7 years, which is lower than Case 2, and advocates how overloading negatively impacts the transformer lifetime. It is interesting to see that this considerable decrease in transformer lifetime is a result of a limited overload in a limited number of hours, which shows the significant impact of overloading on transformer lifetime.

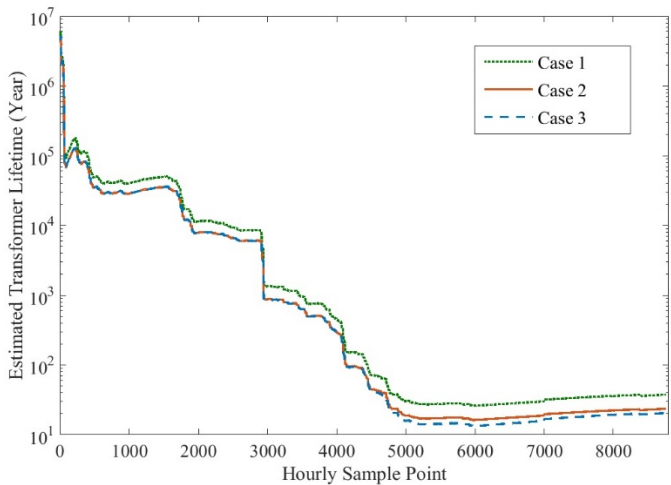

Fig. 7. Comparison of estimating transformer lifetime up to its convergence in all studied cases.

## IV. Conclusions

Power transformers, which are not only considered as a mainstay of providing reliable power to customers, but are also expensive relative to other power system components, have always played a major role in asset management. In this paper, by leveraging sensory data, an efficient approach in estimating transformer lifetime was proposed. Measuring the hourly winding hottest-spot temperature via the temperature sensor, and employing the CMA model to the data stream of the transformer loss of life, the transformer lifetime was estimated at each hour, until it was converged to a constant value. Comparing this calculated lifetime with the time that the transformer has been in service, would provide the remaining lifetime of the asset. The proposed approach was analyzed through numerical simulations under different weather conditions and transformer's loading, where it was shown that overloading could potentially lead to significant drop in the transformer lifetime. Utility companies can reap the benefits of this simple, practical, and yet intelligent approach for transformer asset management, without the need for additional investment in the system.